\documentclass[allclo]{FBSart}
\usepackage{amsfonts}
\usepackage{amssymb}
\input psfig.sty

\title{Charge-dependence of the $\pi NN$ coupling constant and
charge-dependence of the $NN$ interaction\thanks{Dedicated to 
Walter Gl\"ockle on the occasion of his 60th birthday.}}
\author{R. Machleidt$^a$\thanks{E-mail address:
machleid@uidaho.edu} and
        M. K. Banerjee$^b$\thanks{E-mail address: manojb@physics.umd.edu}}
\institute{$^a$Department of Physics, University of Idaho, Moscow,
Idaho 83844, U. S. A.\\
        $^b$Department of Physics, University of Maryland, College Park,
Maryland 20742, U. S. A.}

\runningauthor{R.\ Machleidt and M.\ K.\ Banerjee}
\runningtitle{CD of $\pi NN$ coupling constant and CD of $NN$ interaction}
\begin{document}

\maketitle

\begin{abstract}
The recent determination of the charged $\pi NN$ coupling constant,
$g_{\pi^\pm}$, 
by the Uppsala Neutron Research Group
implies that there may be considerable charge-splitting
of the pion coupling constant.
We investigate the consequences of this for the 
charge-independence breaking (CIB) of the 
$^1S_0$ scattering length, 
$\Delta a_{CIB}$. 
We find that
$\Delta a_{CIB}$ 
depends sensitively on the difference between 
$g_{\pi^\pm}$
and the neutral $\pi NN$ coupling constant,
$g_{\pi^0}$.
Moreover, if
$g^2_{\pi^\pm}$
is only about 3\% larger than
$g^2_{\pi^0}$, then
the established theoretical explanation of
$\Delta a_{CIB}$ 
(in terms of pion mass splitting)
is completely wiped out.
\end{abstract}

\section{Introduction}

From 1973 to 1987, there was a consensus that the $\pi NN$
coupling constant is $g^2_\pi/4\pi=14.3\pm 0.2$
(equivalent to
$f^2_{\pi}=0.079 \pm 0.001$~\cite{foot}). 
This value was obtained by Bugg {\it et al.}~\cite{BCC73}
from the analysis of $\pi^\pm p$ data in 1973, and
confirmed by
Koch and Pietarinen~\cite{KP80} in 1980.
Around that same time,
the neutral-pion coupling constant
was determined by Kroll~\cite{Kro81}
from the analysis of $pp$ data by means of forward dispersion
relations; he obtained
$g^2_{\pi^0}/4\pi = 14.52 \pm 0.40$ 
(equivalent to
$f^2_{\pi^0} = 0.080 \pm 0.002$).

The picture changed in 1987, when the Nijmegen group~\cite{Ber87}
determined the neutral-pion coupling constant in a partial-wave analysis
of $pp$ data and obtained 
$g^2_{\pi^0}/4\pi = 13.1 \pm 0.1$.
Including also the magnetic moment interaction between protons in the analysis,
the value shifted to $13.55 \pm 0.13$ in 1990~\cite{Ber90}.
Triggered by these events, Arndt {\it et al.}~\cite{Arn90} reanalysed
the $\pi^\pm p$ data to determine the charged-pion coupling constant
and obtained 
$g^2_{\pi^\pm}/4\pi = 13.31 \pm 0.27$.
In subsequent work, the Nijmegen group also analysed $np$, $\bar{p}p$,
and $\pi N$ data. The status of their work as of 1993 is summarized in
Ref.~\cite{STS93} where they claim that the most accurate values 
are obtained in their combined $pp$ and $np$ analysis yielding
$g^2_{\pi^0}/4\pi = 13.47 \pm 0.11$ 
(equivalent to 
$f^2_{\pi^0} = 0.0745 \pm 0.0006$)
and
$g^2_{\pi^\pm}/4\pi = 13.54 \pm 0.05$ 
(equivalent to
$f^2_{\pi^\pm}=0.0748 \pm 0.0003$).
The latest analysis of all $\pi^\pm p$ data below 2.1 GeV conducted by
the VPI group
using fixed-$t$ and forward dispersion relation constraints
has generated
$g^2_{\pi^\pm}/4\pi=13.75\pm 0.15$~\cite{AWP94}.
The VPI $NN$ analysis extracted
$g^2_{\pi^0}/4\pi \approx 13.3$ and 
$g^2_{\pi^\pm}/4\pi \approx 13.9$ as well as
the charge-independent value
$g^2_{\pi}/4\pi \approx 13.7$  
\cite{ASW94,ASW95}.

Also Bugg and coworkers
have performed new determinations of the $\pi NN$ coupling
constant. Based upon precise $\pi^\pm p$ data in the 100--310 MeV range
and applying fixed-$t$ dispersion relations, they obtained the value
$g^2_{\pi^\pm}/4\pi=13.96\pm 0.25$
(equivalent to 
$f^2_{\pi^\pm} = 0.0771 \pm 0.0014$)~\cite{MB93}.
From the analysis of $NN$ elastic data between
210 and 800 MeV,
Bugg and Machleidt~\cite{BM95} have deduced
$g^2_{\pi^\pm}/4\pi = 13.69\pm 0.39$ and
$g^2_{\pi^0}/4\pi = 13.94\pm 0.24$.

Thus, it may appear that recent determinations show a consistent
trend towards a lower value for $g_\pi$ with no indication for
substantial charge dependence.

Unfortunately this is not true. There is
one recent determination that does
not follow the current trend.
Using the Chew extrapolation procedure, 
the Uppsala Neutron Research Group has deduced 
the charged-pion coupling constant from high
precision $np$ charge-exchange data at 162 MeV~\cite{Eri95}.
Their latest result is
$g^2_{\pi^\pm}/4\pi = 14.52\pm 0.26$~\cite{Rah98}.
We note that the method used by the Uppsala Group 
is controversial~\cite{ASW95,SRT98}.

If one tries to summarize the confusing current picture
then one may state that recent determinations of the neutral-pion 
coupling constant are, indeed, consistently on the low side 
with a value of 
$g^2_{\pi^0}/4\pi = 13.6 \pm 0.3$ 
covering about 
the range of current determinations.

However, there is no such consistent picture for the charged-pion
coupling constant with recent determinations being up to
nine standard deviations apart.
If we trust the Uppsala result of
$g^2_{\pi^\pm}/4\pi = 14.52\pm 0.26$,
then large charge-splitting of $g_\pi$ exists.

This is the motive for the present paper in which we will
investigate the impact of charge-splitting of $g_\pi$ 
on our established theoretical understanding of the charge
dependence of the nuclear force. 
In particular, we will look into the charge-independence breaking (CIB)
of the $^1S_0$ scattering length, $\Delta a_{CIB}$.
We find that
$\Delta a_{CIB}$ 
depends sensitively on the difference between 
$g_{\pi^\pm}$
and 
$g_{\pi^0}$.
Moreover, if
$g_{\pi^\pm}$
is only moderately larger than
$g_{\pi^0}$,
the established theoretical explanation of
$\Delta a_{CIB}$ 
(in terms of pion mass splitting)
is completely wiped out.

\section{Conventional explanation of the charge-dependence
of the $NN$ interaction}
The equality between 
proton-proton ($pp$) [or neutron-neutron ($nn$)] and neutron-proton ($np$)
nuclear interactions is known as charge independence---a symmetry that
is slightly broken.  This is seen most clearly
in the $^1S_0$ nucleon-nucleon scattering lengths. 
The latest empirical values for the singlet scattering length $a$ 
and effective range $r$ are~\cite{MNS90,MO95}:
\begin{equation}
\begin{array}{lll}
a^N_{pp}=-17.3\pm 0.4 \mbox{ fm}, &\hspace*{2.5cm}
                                     & r^N_{pp}=2.85\pm 0.04 \mbox{ fm},\\
a^N_{nn}=-18.8\pm 0.3 \mbox{ fm}, && r^N_{nn} = 2.75\pm 0.11 \mbox{ fm},\\
a_{np}=-23.75\pm 0.01 \mbox{ fm}, && r_{np}=2.75\pm 0.05 \mbox{ fm}.
\end{array}
\end{equation}
The values given for $pp$ and $nn$ 
scattering refer to the nuclear part of the interaction
as indicated by the superscript $N$.
Electromagnetic effects have been removed from the experimental
values, which is model dependent. The uncertainties
quoted for $a^N_{pp}$ and $r^N_{pp}$ are due to this model dependence.

It is useful to define the following averages:
\begin{eqnarray}
\bar{a}\equiv \frac12 (a^N_{pp} + a^N_{nn}) & =&  -18.05\pm 0.5 \mbox{ fm},\\
\bar{r}\equiv \frac12 (r^N_{pp} + r^N_{nn}) & =&  2.80\pm 0.12 \mbox{ fm}.
\end{eqnarray}
By definition, charge-independence breaking (CIB) is the difference between 
the average of $pp$ and $nn$, on the one hand, and $np$ on the other: 
\begin{eqnarray}
\Delta a_{CIB} 
\equiv
 \bar{a}
 - 
 a_{np}
 &=& 5.7\pm 0.5 \mbox{ fm},\\
\Delta r_{CIB} \equiv
 \bar{r}
 - 
 r_{np}
 &=& 0.05\pm 0.13 \mbox{ fm}.
\end{eqnarray}
Thus, the $NN$ singlet scattering length shows a clear signature
of CIB in strong interactions.

The current understanding is 
that the charge dependence of nuclear forces is due to
differences in the up and down quark masses and electromagnetic 
interactions. 
On a more phenomenological level, major causes of CIB are the
mass splittings of isovector mesons (particularly, $\pi$ and $\rho$)
and irreducible pion-photon exchanges.

It has been known for a long time that the difference between the charged and 
neutral pion masses in the one-pion-exchange (OPE) potential  accounts
for about 50\% of $\Delta a_{CIB}$. 
Based upon the Bonn meson-exchange model for the $NN$ 
interaction~\cite{MHE87}, also multiple pion exchanges have been taken
into account. Including these interactions, about
80\% of the empirical $\Delta a_{CIB}$ can be explained~\cite{CM86,LM98}.
Ericson and Miller~\cite{EM83} obtained a very similar result using the
meson-exchange model of Partovi and Lomon~\cite{PL70}.

The CIB effect from OPE can be understood as follows.
In nonrelativistic approximation~\cite{foot2} and disregarding isospin
factors, OPE is given by
\begin{equation}
V_{1\pi}(g_\pi, m_\pi)  =  -\frac{g_{\pi}^{2}}{4M^{2}}
 \frac{({\mbox {\boldmath $\sigma$}}_{1} \cdot {\bf k})
       ({\mbox {\boldmath $\sigma$}}_{2} \cdot {\bf k})}
{m_{\pi}^2+{\bf k}^{2}}
\left( \frac{\Lambda^{2}-m_{\pi}^{2}}{\Lambda^{2}+{\bf k}^{2}}
\right)^n
\end{equation}
with $M$ the average nucleon mass, $m_\pi$ the pion mass, 
and {\bf k} the momentum transfer.
The above expression includes a form factor with 
cutoff mass $\Lambda$ and exponent $n$.

For $S=0$ and $T=1$, where $S$ and $T$ denote the total spin and isospin
of the two-nucleon system,
respectively, we have
\begin{equation}
V_{1\pi}^{01}(g_\pi, m_\pi)  =  
\frac{g_{\pi}^{2}}
{m_{\pi}^2+{\bf k}^{2}}
\frac{{\bf k}^2}
{4M^{2}}
\left( \frac{\Lambda^{2}-m_{\pi}^{2}}{\Lambda^{2}+{\bf k}^{2}}
\right)^n \; ,
\end{equation}
where the superscripts 01 refer to $ST$.
In the $^1S_0$ state, this potential expression is repulsive.
The charge-dependent OPE is then,
\begin{equation}
V_{1\pi}^{01}(pp)  =  
V_{1\pi}^{01}(g_{\pi^0}, m_{\pi^0})  
\end{equation}
for $pp$ scattering, and
\begin{equation}
V_{1\pi}^{01}(np)  =  
2 V_{1\pi}^{01}(g_{\pi^\pm}, m_{\pi^\pm})  
- V_{1\pi}^{01}(g_{\pi^0}, m_{\pi^0})  
\end{equation}
for $np$ scattering.

If we assume charge-independence of $g_\pi$ (i.~e., 
$g_{\pi^0}=g_{\pi^\pm}$), then all CIB comes from the charge
splitting of the pion mass, which is~\cite{PDG96}
\begin{eqnarray}
m_{\pi^0} & = & 134.976 \mbox{MeV,}\\
m_{\pi^\pm} & = & 139.570 \mbox{MeV.}
\end{eqnarray}

\begin{table}
\caption{Predictions for $\Delta a_{CIB}$ as defined in Eq.~(4)
in units of fm without and with the assumption of charge-dependence
of $g_\pi$.}
\small
\begin{tabular}{cccc}
\hline
\hline
 & \multicolumn{2}{c}{\bf No charge-dependence of $g_\pi$} 
                                     & {\bf Charge-dependent $g_\pi$:} \\ 
 &                    &                           &$g^2_{\pi^0}/4\pi = 13.6$\\
 &Ericson \& Miller~\cite{EM83}&Li \& Machleidt~\cite{LM98}
                                            &$g^2_{\pi^\pm}/4\pi = 14.4$\\
\hline
$1\pi$ & 3.50 & 3.24 & --1.58 \\
$2\pi$ & 0.88 & 0.36 & --1.94 \\
$\pi\rho,\pi\sigma,\pi\omega$ & --- & 1.04 & --0.97 \\
\hline
Sum & 4.38 & 4.64 & --4.49 \\
\hline
Empirical & \multicolumn{3}{c}{$5.7\pm 0.5$}
\\
\hline
\hline
\end{tabular}
\end{table}

Since the pion mass appears in the denominator of OPE,
the smaller $\pi^0$-mass exchanged in $pp$ scattering
generates a larger (repulsive) potential in the $^1S_0$
state as compared to $np$ where also the larger $\pi^\pm$-mass
is involved. Moreover, the $\pi^0$-exchange in $np$
scattering carries a negative sign, 
which further weakens the $np$ OPE potential.
The bottom line is that the $pp$ potential is more repulsive
than the $np$ potential. The quantitative effect on
$\Delta a_{CIB}$
is about 3 fm (cf.\ Table 1).

\begin{figure}
\psfig{file=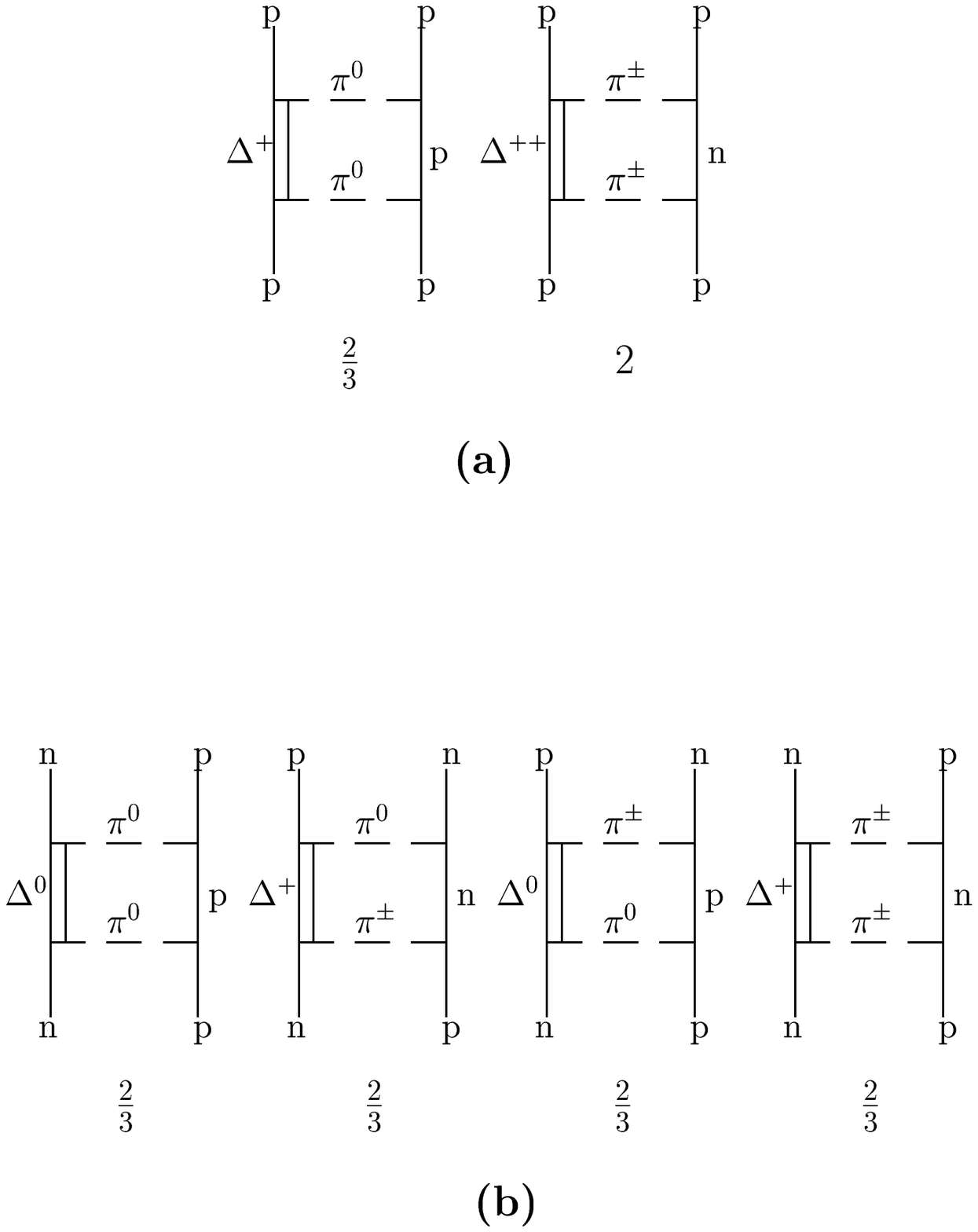,width=12cm}
\caption{$2\pi$-exchange box diagrams
with $N\Delta$ intermediate states that contribute to 
(a) $pp$ and (b) $np$ scattering. 
The numbers below the diagrams are
the isospin factors.}
\end{figure}

We now turn to the CIB created by the $2\pi$ exchange (TPE) contribution
to the $NN$ interaction. There are many diagrams that
contribute (see Ref.~\cite{LM98} for a complete overview).
For our qualitative discussion here, we pick the largest
of all $2\pi$ diagrams, namely, the box diagrams with
$N\Delta$ intermediate states, Fig.~1.
Disregarding isospin factors and using some drastic
approximations~\cite{foot2}, the amplitude for such a diagram is
\begin{equation}
V_{2\pi}(g_\pi, m_\pi)  =  -\frac{g_{\pi}^{4}}{16M^{4}}
\frac{72}{25} \int \frac{d^3p}{(2\pi)^3}
 \frac{[{\mbox {\boldmath $\sigma$}} \cdot {\bf k}
        {\mbox {\boldmath $S$}} \cdot {\bf k}]^2}
{(m_{\pi}^2+{\bf k}^{2})^2(E_p+E^\Delta_p-2E_q)}
\left( \frac{\Lambda^{2}-m_{\pi}^{2}}{\Lambda^{2}+{\bf k}^{2}}
\right)^{2n} \; ,
\end{equation}
where ${\bf k} = {\bf p} - {\bf q}$ with {\bf q}
the relative momentum in the initial and final state
(for simplicity, we are considering a diagonal matrix element); 
$E_p=\sqrt{M^2+{\bf p}^2}$ and $E^\Delta_p=\sqrt{M_\Delta^2+{\bf p}^2}$
with $M_\Delta=1232$ MeV the $\Delta$-isobar mass;
{\bf S} is the spin transition operator between nucleon and
$\Delta$. For the $\pi N\Delta$ coupling constant, $f_{\pi N\Delta}$,
the quark-model relationship
$f^2_{\pi N\Delta} = \frac{72}{25} f^2_{\pi NN}$ 
is used~\cite{MHE87}.

For small momentum transfers {\bf k},
this attractive contribution is roughly proportional to
$m_\pi^{-4}$. Thus for TPE, the heavier pions will provide less attraction
than the lighter ones.
Charged and neutral pion exchanges
occur for $pp$ as well as for $np$, and it is important
to take the isospin factors carried by the various diagrams
into account. They are given in Fig.~1 below each diagram.
For $pp$ scattering, the diagram with 
double $\pi^\pm$ exchange
carries the largest factor, while 
double $\pi^\pm$ exchange
carries only a small relative weight in $np$ scattering.
Consequently, $pp$ scattering is less attractive than $np$
scattering which leads to an increase of
$\Delta a_{CIB}$ by 0.79 fm due to the diagrams of Fig.~1. 
The crossed diagrams of this type
reduce this result and including all $2\pi$ exchange diagrams 
one finds a total effect of 0.36 fm~\cite{LM98}.
Diagrams that go beyond $2\pi$ have also been investigated
and contribute another 1 fm (see Table 1 for a summary).

In this way, pion-mass splitting explains about 80\% of 
$\Delta a_{CIB}$.

\section{Charge-dependence of the pion coupling constant and 
charge-dependence of the singlet scattering length}

In this section, we will consider also charge-splitting of $g_\pi$,
besides pion mass splitting.

As discussed in the Introduction, some current determinations of $g_\pi$
may suggest the values
\begin{eqnarray}
g^2_{\pi^0}/4\pi & = & 13.6 \; , \\
g^2_{\pi^\pm}/4\pi & = & 14.4 \: .
\end{eqnarray}
Accidentally, this splitting is---in relative terms---about the same as the
pion-mass splitting; that is
\begin{equation}
\frac{g_{\pi^0}}{m_{\pi^0}} \approx
\frac{g_{\pi^\pm}}{m_{\pi^\pm}}\; . 
\end{equation}
From the discussion in the previous section, we know
that (for zero momentum transfer)
\begin{equation}
\mbox{OPE} \sim \left(\frac{g_\pi}{m_\pi}\right)^2
\end{equation}
and
\begin{equation}
\mbox{TPE} \sim \left(\frac{g_\pi}{m_\pi}\right)^4 \; ,
\end{equation}
which is not unexpected, anyhow.
On the level of this qualitative discussion, we can then predict that
any pionic charge-splitting 
satisfying Eq.~(15) will create no CIB from pion exchanges.
Consequently, a charge-splitting of $g_\pi$ as given in Eqs.~(13)
and (14) will wipe out our established explanation of CIB
of the $NN$ interaction.

We have also conducted accurate numerical calculations based upon the Bonn
meson-exchange model for the $NN$ interaction~\cite{MHE87}.
The details of these calculations are spelled out
in Ref.~\cite{LM98} where, however, no charge-splitting of $g_\pi$
was considered. Assuming the $g_\pi$ of 
Eqs.~(13) and (14), we obtain the $\Delta a_{CIB}$ 
predictions given in the last column of Table~1.
It is seen that the results of an accurate calculation go even beyond
what the qualitative estimate suggested:
the conventional CIB prediction is not only reduced, it is reversed.
This is easily understood if one recalls that the pion mass appears
in the propagator $(m_\pi^2+{\bf k}^2)^{-1}$. Assuming an
average ${\bf k}^2\approx m^2_\pi$, the 7\% charge splitting of
$m^2_\pi$ will lead to only about a 3\% charge-dependent effect from
the propagator. Thus, if a 6\% charge-splitting of $g_\pi^2$ is used, 
this will not only override the pion-mass effect, it will reverse it.

Based upon this argument and on our numerical results, 
one can then estimate that
a charge-splitting of $g_\pi^2$ of only about 3\%
(e.~g., 
$g^2_{\pi^0}/4\pi = 13.6$ and $g^2_{\pi^\pm}/4\pi = 14.0$)
would erase all CIB prediction of the singlet scattering length
that is based upon the conventional mechanism of pion mass splitting.

\section{Conclusions}

All current determinations of the neutral-pion coupling constant
seem to agree on a `low' value, like
$g^2_{\pi^0}/4\pi 
= 13.6 \pm 0.3$. 
However, for the charged-pion coupling constant, there is no
such agreement.
While some recent determinations of 
$g^2_{\pi^\pm}/4\pi$ 
come up with a value close to
$g^2_{\pi^0}/4\pi$, 
the Uppsala group~\cite{Rah98}
obtains
$g^2_{\pi^\pm}/4\pi = 14.52\pm 0.26$
which implies a large charge-dependence of $g_\pi$.

In this paper, we have investigated the consequences of such a large
charge-dependence of $g_\pi$ for the conventional explanation of the
charge-dependence of the $^1S_0$ scattering length, $a_s$.
We find that a charge-splitting of the coupling constant, defined by
$\Delta g^2_\pi/4\pi \equiv (g^2_{\pi^\pm} - g^2_{\pi^0})/4\pi$,
of $\Delta g^2_\pi/4\pi = 0.4$ 
would wipe out the effect of the conventional mechanism
(namely, pion mass splitting) and a splitting of
$\Delta g^2_\pi/4\pi = 0.8$ 
would even reverse the charge-dependence of $a_s$~\cite{foot3}.

Besides pion mass splitting, we do not know of any other essential mechanism
to explain the charge-dependence of $a_s$. Therefore, it is unlikely
that this mechnism is annihilated by a charge-splitting of $g_\pi$.
This may be taken as an indication that there is no significant
charge splitting of the $\pi NN$ coupling constant.

Consequently, charge-dependence of $g_\pi$ is most likely not
the resolution of the large differences in recent $g_\pi$
determinations; which implies that we are dealing here
with true discrepancies. The reasons for these discrepancies
may be large (unknown) sytematic errors and/or a gross
underestimation of the errors in essentially all present $g_\pi$
determinations.

\begin{acknowledge}
This work was supported in part by the U.S. National Science Foundation
under Grant-No.\ PHY-9603097 and by the U.S. Department of Energy
under Contract No.\ DE-FG02-936R-40762.
\end{acknowledge}


\begin{thebibliography}{99}
\bibitem{foot2} For pedagogical reasons, we use simple, approximate
expressions when discussing the effects from pion exchange in a
qualitative manner. Note, however, that in our numerical
calculations we use relativistic time-ordered perturbation
theory in its full complexity and without approximations,
as outlined in all detail in the appendices of Ref.~\cite{MHE87}.
\bibitem{foot3} We note that a negative value for
$\Delta g^2_\pi/4\pi$
would enhance
$\Delta a_{CIB}$.
Since pion mass splitting explains only 80\% of
$\Delta a_{CIB}$,
a charge-dependent coupling constant splitting of
$\Delta g^2_\pi/4\pi \approx -0.1$
(e.g.,
$g^2_{\pi^0}/4\pi = 13.6$ and
$g^2_{\pi^\pm}/4\pi = 13.5$)
would explain the remaining 20\%.
\end{thebibliography}
\end{document}